\documentclass[12pt,preprint]{aastex}

\slugcomment{Accepted by Astronomical Journal.  Expected publication
  date: September 2005.}

\shorttitle{The Cornell High-order Adaptive Optics Survey for Brown
  Dwarfs in Stellar Systems--I:  Observations, Data Reduction, and
  Detection Analyses}
\shortauthors{Carson et al.}

\begin{document}

\title{The Cornell High-order Adaptive Optics Survey for Brown Dwarfs
  in Stellar Systems--I:  Observations, Data Reduction, and Detection Analyses}

\author{J. C. Carson\altaffilmark{1,2}, S. S. Eikenberry\altaffilmark{3},
  B. R. Brandl\altaffilmark{4}, J. C. Wilson\altaffilmark{5}, and T. L. Hayward\altaffilmark{6}}

\affil{Department of Astronomy, Cornell University, Ithaca, NY 14853}

\altaffiltext{1}{present address: Jet Propulsion Laboratory, Earth \&
  Space Sciences, 4800 Oak Grove Dr., MS 183-900, Pasadena, CA 91109}
\altaffiltext{2}{Postdoctoral Scholar, California Institute of Technology.}
\altaffiltext{3}{present address: Department of Astronomy, University of Florida, 211 Bryant
  Space Science Center, Gainesville, FL 32611}
\altaffiltext{4}{present address: Leiden Observatory, PO Box 9513,
  2300 RA Leiden, The Netherlands}
\altaffiltext{5}{present address: Department of Astronomy, University of Virginia, 255 Astronomy Building, PO Box 3818, Charlottesville, VA 22903} 
\altaffiltext{6}{present address: Gemini Observatory, Southern Operations Center, Casilla 603 La Serena Chile}

\begin{abstract}
     In this first of a two-paper sequence, we report techniques and results of the Cornell High-order
     Adaptive Optics Survey for brown dwarf companions (CHAOS).
     At the time of this writing, this study represents the most
     sensitive published population survey of brown dwarf companions to main
     sequence stars, for separation akin to our own outer solar system.
     The survey, conducted using the Palomar 200-inch
     Hale Telescope, consists of K$_{s}$ coronagraphic observations of 80
     main sequence stars out to 22 parsecs.  At 1$\arcsec$ separations from a typical target
     system, the survey achieves median sensitivities 10 magnitudes
     fainter than the parent star.  In terms of companion mass,
     the survey achieves typical sensitivities of 25
     M$_{Jup}$ (1 Gyr), 50 M$_{Jup}$ (solar age), and 60 M$_{Jup}$ (10
     Gyr), using evolutionary models of Baraffe et al. (2003).  Using
     common proper motion to distinguish companions from field stars,
     we find that no systems show positive evidence of a substellar
     companion (searchable separation $\sim$ 1-15$\arcsec$
     [projected separation $\sim$ 10-155 AU at the median target distance]).  In the second paper of the series we shall
     present our Monte Carlo population simulations.       
\end{abstract}


\section{Introduction}
    The discovery of the brown dwarf Gl 229B (Nakajima et al. 1995)
    heralded a stream of direct detections of sub-stellar objects.
    Field surveys such as the Two Micron All-Sky Survey (2MASS;
    Skrutskie et al. 1997), the Deep Near Infrared Survey (DENIS;
    Epchtein et al. 1997), and the Sloan Digital Sky Survey (SDSS;
    Gunn and Weinberg 1995) helped raise the number of brown dwarf
    identifications today to close to a thousand.  But despite these advances, the search for
    brown dwarf companions at intermediate and narrow separations (say
    less than a few arcseconds) to main sequence stars remains difficult.  Despite a strong
    community effort in
    high-contrast imaging observations, less than a half-dozen companions have been
    confirmed as $<$100 AU (projected separation) substellar
    companions to main sequence stars.  

    To date, the most comprehensive probe of ultra-narrow
     separation ($\lesssim$10 AU) brown dwarf
     companions comes from radial velocity surveys such as Marcy \&
     Butler (2000) :  McCarthy \& Zuckerman (2004), for instance, report that over 1500 F,
     G, K, and M stars have been observed, via this method, with sensitivities strong
     enough to detect brown dwarf companions between 0 and 5 AU.
     Using observations like these, Marcy \& Butler (2000) conclude
     that the
     $\leq$3 AU brown dwarf companion fraction to F-M stars is less
     than 0.5$\%$.  High angular resolution imaging surveys such as
     CHAOS, a deep adaptive optics (AO) coronagraphic search and proper motion follow-up of faint companions, are required to test if this low companion fraction extends out
     to intermediate distances, akin to our own outer solar sytem.
     Knowledge of these intermediate separation companions will help
     bridge the gap between radial velocity companion surveys and
     wide-separation companion data such as 2MASS.  The CHAOS survey is
     not the first study to examine this search space.  Recently
     published intermediate separation coronagraphic surveys include Liu et
     al. (2002), Luhman \& Jayawardhana (2002), McCarthy \& Zuckerman
     (2004), Metchev \& Hillenbrand (2004), and Potter et al. (2002).
     Unlike these other surveys however, the CHAOS paper here represents, at
     the time of this writing, the only published adaptive
     optics survey that reports complete results for all surveyed
     targets, including null result observations.  Reporting
     comprehensive results on all surveyed targets, this CHAOS paper
     invites a rich opportunity for statistical inquiry into brown
     dwarf and stellar formation theory.

     In the sections below we present techniques and results for the recently
     completed CHAOS survey.  Section 2 presents our target
     sample.  Section 3 describes our observing techniques.
     In Section 4 we present the data analysis techniques we developed for this
     survey.  In
     Section
     5 we summarize our survey sensitivities.  Section 6
     describes our results.  We present our
     conclusions in Section 7.

\section{Target Sample}
   We began our candidate selection process with a careful review of
   the Third Catalogue of Nearby Stars (Gliese \& Jahreiss 1995).
   Beginning with northern stars, we prioritized targets by their
   closeness to our solar system.  Next we discarded all stars that
   exist in known resolvable multiple systems, as this scenario would prevent us
   from effectively hiding the entire parent system behind the
   0.$\arcsec$9 coronagraphic mask.  We double-checked for the
   presence of stellar companions using Hipparcos data (Perryman et
   al. 1997) as well as on-telescope preliminary imaging.  The
   one known (Perryman et al. 1997) resolvable binary that we kept was
   Gliese 572;  For this target, the
   secondary star's narrow separation ($\sim$0.$\arcsec$4)
   allowed us to hide both stars behind the
   0.$\arcsec$9 coronagraphic spot, to an acceptable level.  We did
   not delete spectroscopic binaries from the target list as these unresolved targets
   allowed for effective coronagraphic masking;  Gliese 848, 92,
   567, 678, and 688 of our final target list are known spectroscopic
   binaries, as published in Pourbaix et al. (2004).     As the next
   step, we removed all stars with a V magnitude fainter than $\sim$
   12 mags.  Our previous experience using Palomar Adaptive Optics
   (PALAO) in 2000 indicated that stars fainter than this limit were unable to
   serve as effective natural guide stars.  Next we searched the
   USNO-A2.0 Catalogue (Monet et al. 1998) for a corresponding point
   spread function (PSF) calibration star for each targeted star.  For
   choosing a PSF calibration star, we required the following
   restrictions:  1)  A separation less than a couple degrees from the
   target star;  2)  A difference in V magnitude, relative to the
   target star, $\lesssim$ 1 mag;  3)  An absence of any known
   resolvable companions.  These restrictions ensured that the calibration star
   would deliver a measured PSF similar to the target star's PSF.  Any
   target star that did not have a corresponding calibration star
   meeting this criterion was removed from the sample.  We expanded
   the search region further and further south from the original
   northern positions until the list included a total of 80 stars
   extending as far south as -10 degrees inclination.  This final
   target sample included 3 A stars, 8 F stars, 13 G stars, 29 K
   stars, 25 M stars, and 2 stars with ambiguous spectral types.  
   All stars possessed well-characterized proper motion values as
   defined by Hipparcos (Perryman et al. 1997).  As nearby stars, they
   typically possessed high proper motion (median target proper motion
   $\sim$ 600 mas/yr)
   thus facilitating an efficient common proper motion follow-up
   strategy for candidate companions.   A complete list of the target set is given in Table 1.
   
        In the second paper of this series, we will present a
   thorough discussion of how selection biases in our sample may
   affect derived brown dwarf populations.  For the time
   being, however, we do
   note that certain formation models, such as ones that support the creation of brown
   dwarfs within multiple systems (Clarke, Reipurth, \&
   Delgado-Donate 2004 for example), imply that observed population
   levels, as derived from our mostly single-star sample, may differ significantly from statistics
   that include multiple
   systems.

\section{Observations}

\subsection[Coronagraphic Search Observations]{Coronagraphic Search Observations}

     To conduct our survey, we used the Palomar Adaptive Optics system
     (PALAO; Troy et al. 2000) and accompanying PHARO science camera
     (Hayward et al. 2001) installed on the Palomar 200-inch Hale
     Telescope.  PALAO provided us with the high resolution (FWHM
     typically $\sim$ 0$\arcsec$.14 in K-short) necessary for
     resolving close companions.  The accompanying PHARO science
     camera       (wavelength sensitivity 1-2.5 $\micron$ and
     platescale 40 mas per pixel)  provided us with a coronagraphic imaging capability along
     with a field of view ($\sim$30$\arcsec$) substantially larger
     than any competitively sized telescope's adaptive optics system, at the time of the
     survey's commencement. 

     Our general observing strategy was to align the coronagraphic
     mask on a target star and take a series of short exposures as to
     not saturate many pixels in the detector.  (Occasionally we
     saturated at the edges of the coronagraphic mask where high noise
     levels already prevented any meaningful companion search.)  We
     planned our exposure time and number of exposures to allow for a maximum 8 minutes of
     execution time (including overheads).  This helped  ensure that sky conditions did not significantly
     change between the target exposures and following PSF calibration
     star exposures.  When target star exposures were complete, we
     spent a similar amount of time taking coronagraphic images of
     the
     PSF calibration star.
     Immediately flanking this target pair, we took dithered images of
     a nearby empty sky region, using the same set-up as the target
     and reference star series.  We repeated this process (sky,
     target, reference star, sky) as many
     times as necessary to reach our desired signal to noise.  Once we
     completed these image sets, we inserted a neutral density filter
     in the optical path and conducted dithered non-coronagraph exposures of
     the target star.  These images allowed us to characterize and
     record instrument and site observing conditions.  Table 1
     describes relevant observing information for the individual targets. 

\subsection{Common Proper Motion Observations}

     For candidate companions detected in the previous procedures, we
     checked for a physical companionship by using common proper
     motion observations.  The nearby stars we observed tend to have
     high proper motions (on the order of a few hundred
     mas yr$^{-1}$).   The vast majority of false candidate companions are
     background stars that tend to have very small proper motions
     compared to the parent star.  Therefore, after recording our
     initial measurement, we waited for a timespan long enough for the
     parent star to move a detectable distance, typically
     $\gtrsim$3 sigma separation from the original position.  We then repeated our
     observing set so that we could check to see if the candidate
     maintained the same position with respect to the parent star.
     Target stars re-observed to check for common proper motion include
     Gliese 740, 75, 172, 124, 69, 892, 752, 673, 41, 349, 412, 451,
     390, 678, 768, 809, 49, and 688.  Due to instrument scheduling
     constraints, Gliese 49, 41, 390, and 678 were all re-observed
     using the Palomar 200-inch Wide-field Infrared Camera (WIRC;
     Wilson et al. 2003) rather than PALAO and the accompanying PHARO
     science camera.  Since the WIRC camera possesses no coronagraphic
     mode, the observations were instead conducted using standard
     dithered exposure sequences.  WIRC, with its
     non-AO-corrected point spread function and lack of a
     coronagraphic mask, made a poorer probe of astrometry than the
     PHARO camera.  However, the systems observed with WIRC all
     possessed large expected proper motions ($>$400 mas [1 WIRC pixel
     $\sim$250 mas]) and large separations ($>$10 arcseconds) from the
     parent system, making them acceptable WIRC observing targets. 

\section{Data Analysis}

\subsection{Reducing Images}

     We began our data reduction by median-combining each of the
     dithered sky sets.  We then took each coronagraphed star image
     and subtracted the median-combined sky taken closest in time to
     the star image.  (The typical separation in time between target
     and sky image was $\sim$5.5 minutes.)  We divided each of the
     sky-subtracted star images by a flatfield frame that we created,
     using standard procedures, from twilight calibration images taken
     that same night.  Next we median-combined each sequence of
     coronagraphed star frames.  For this median-combination, we used
     the images' residual parent star flux (that leaked from around
     the coronagraph) to realign any frames that may have shifted due
     to instrument flexure.  Next we applied a bad pixel algorithm to
     remove suspicious pixels (defined as any pixel deviating
     from the surrounding 8 pixels by $\geq$ 5-sigma)
     and replace them with the median of their neighbors.

     After completing this procedure for both target star and
     calibration star image sets, we scaled the calibration star PSF
     so that two 50-90 pixel annuli, one centered on the target
     star, the other centered on the scaled PSF, exhibited identical median
     values.  Next we multiplied the scaled PSF by test values ranging
     from 0.20 to 1.76 at 0.04 intervals.  For each test value we also
     tried shifting the scaled PSF -7 to +7 pixels, at
     integar steps, in each of the x and y directions.  From these
     test combinations we selected the adjusted PSF that most closely
     resembled the target star, according to a least-squared
     fit of flux values 50 to 90 pixels from the star center.  We next
     subtracted our adjusted PSF from the target star to arrive at a
     final image for the set.  In the
     cases where we had multiple target star/calibration star
     observing set pairs, we co-added the final images, using the
     residual parent star PSF to correct any misalignments.  

As our final data reduction procedure, we
     applied a Fourier filter to help remove non-point-like features
     such as unwanted internal instrument reflection and residual
     parent star flux.  The Fourier filter application entailed
     our multiplying each pixel in a Fourier transformed version 
     of the final
     image  (where the lowest frequencies resided at the center of the array
     and largest frequencies resided toward the edges) by {\textit
     e$^{\frac{r-23}{34}}$}. {\textit r} here is the separation, in
     units of pixels, between a given pixel and the center of the Fourier
     transformed array.  We then applied an
     inverse Fourier transform to the array to produce the final filtered image.
     We chose the two aforementioned numerical parameters (23 and 34) of our
     exponential function after first running test trials on a crowded field
     image using a generic exponential function 
{\textit e$^{\frac{r-m}{\sigma}}$}. For these trial functions we set
     {\textit m} to test values ranging from 5 to 49, at integar
     values;  We set $\sigma$ to test values ranging from 1 to
     39, again at integar intervals.  Testing all combinations,
     including an equivalent
     gaussian version as well, we found that {\textit
     e$^{\frac{r-23}{34}}$}  produced the greatest signal-to-noise
     improvements.  For sampled field stars in our tests, signal-to-noise levels
     improved by about 25\% between non-filtered and filtered images.
     Along with this signal-to-noise improvement, the typical PSF FWHM
     decreased by about 10\% as a result of the Fourier filter application.

\subsection{Identifying Brown Dwarf Companions}

         Our first step in identifying brown dwarf companions was to
         individually inspect each final psf-subtracted and non-psf-subtracted image for any potential
         companions.  By choosing to examine both subtracted
         and non-subtracted final images, we effectively recognize
         that the PSF-subtraction improves our ability to identify
         candidates close to the parent star, but, due to the
         introduced increased sky noise, makes it more difficult to
         identify candidates at larger separations from the parent star.
         For our identifications, the characteristic Palomar adaptive optics ``waffle
         pattern'' (see Figure 1) helped distinguish real objects from
         false ones.  Practically, we found that this individual
         inspection was the most effective method of identifying
         candidate companions.  However, for the purpose of
         determining quantifiable detection sensitivities, we chose to
         use an automated detection system as well.
          
Our automated algorithm operated by centering on every other
  pixel in the image array
  and creating there a 0.$\arcsec$16 diameter flux aperture and
  1.$\arcsec$2-1.$\arcsec$6 diameter sky annulus.  After subtracting any
  residual sky background, the algorithm approximated a
  signal-to-noise level by dividing the measured aperture flux by the
  combined aperture flux Poisson noise and background noise;  It
  approximated background noise from the standard deviation of the sky annulus pixels. In
          the end it outputted a final array with a signal-to-noise
          value for each sampled pixel.  For each signal-to-noise map,
  it also generated a map of measured background noise at each
  position (as estimated from the sky annuli).  This
  outputted noise map essentially reflected the ability of the
  algorithm to detect (at a given thresh-hold signal-to-noise level)
  different brightness objects according to position on the array.         

After generating maps for a given image, the program selected
          the signal-to-noise map pixel with the highest value, using a
          minimum value of five.  It recorded the
          pixel position and then moved on to record the next highest
          signal to noise value greater than five.  After each
          detection, it voided a 0.$\arcsec$4 radius around the
          detected candidate object.  This procedure continued
          until there were no more positions with signal to noise
          values greater or equal to five.  (Of course, for many
          images, no positions possessed signal to noise levels
          greater than five.)  After the algorithm identified the candidate sources,
          we re-examined the final images to ensure that the
          algorithm had indeed detected a true source as opposed to a
          systematic effect.  Again, we searched for the Palomar adaptive
          optics signature ``waffle pattern'' to ensure a true physical
          source.  We also made comparisons to images taken at other
          sources to ensure that the feature was indeed unique
          to the target image.  

 We acknowledge that
          the use of our automated detection routine has some
          drawbacks.  Notably, there are several instances where the
          algorithm over-estimates the noise level in the
          non-psf-subtracted and psf-subtracted images.  For instance, when
          examining the non-psf-subtracted images at regions near the
          parent star, the algorithm can
          mistake what may be a well-ordered parent star PSF
          slope for a random fluctuation in background noise.  In this
          instance, fortunately, the subsequent examination of the corresponding
          psf-subtracted image should ensure that the initial over-estimation
          in noise does not affect final results.  However, such a correction will not occur when
          the algorithm is hunting around field stars;  If a
          field star happens to fall in the sky annulus, the algorithm will
          determine that region to have
          excessively high background noise.  Thus, only the brightest candidate
          objects would be detected near these field star positions.  
  In Section 5 we discuss how we may generate
          limiting magnitudes and brown dwarf mass limits from these
          algorithm-generated noise maps.

       In cases where we positively  identified a potential brown dwarf companion to a
       parent star, we next estimated its apparent K$_{s}$ magnitude,
       using the non-coronagraphed calibration images of the parent star
       and published 2MASS K-magnitudes (Skrutskie et al. 1997).  Resulting magnitudes are
       displayed in Table 2.  Once we established an apparent K$_{s}$
       magnitude, we derived a corresponding absolute K$_{s}$
       magnitude, assuming the candidate had a distance equal to the
       parent system.  Thanks to observational surveys such as
       Hipparcos (Perryman et al. 1997), all of our parent stars had
       well-defined parallaxes and therefore distances.  With an
       approximate absolute K$_{s}$ magnitude in hand, we combined
       published brown dwarf observational data (Leggett et al. 2000,
       Leggett et al. 2002, Burgasser et al. 1999, Burgasser et
       al. 2000,  Burgasser et al. 2002, Burgasser, McElwain, \&
       Kirkpatrick 2003, Geballe et al. 2002, Zapatero et al. 2002,
       Cuby et al. 1999, Tsvetanov et al. 2000, Strauss et al. 1999,
       and Nakajima et al. 1995)  with theoretical data from Burrows
       et al. (2001) to extrapolate constraints on the object's mass.  An object whose potential mass fell within acceptable brown dwarf restrictions was designated for common proper motion follow-up observations.  

     For our follow-up observations, we used Hipparcos published
     common proper motion values (Hipparcos catalogue; Perryman et
     al. 1997) to determine the expected movement of the parent
     system.  Since background and field stars are unlikely to possess
     proper motions identical to the parent system's, we used common
     proper motion as a strong support for a physical companionship.
     To determine the candidate companion's relative position in
     different epoch images, we fit a gaussian profile to the
     candidate companion flux position.  For the parent star, we
     determined position from an extrapolated gaussian profile created
     from the flux leaking from behind the coronagraphic mask.
     We could typically constrain the parent star position to
     within a pixel or two and the candidate position to a fraction of a
     pixel, depending on the signal-to-noise levels. 
     Measuring the candidate companion's relative position over the
     two epochs, we were able to distinguish physical companionships
     from chance alignments.  We record positions in
     Table 2.

\section{ Survey Sensitivities}

\subsection{Determining Limiting Magnitudes}

     To quantify detection sensitivities from the
     algorithm-generated noise maps described
     in Section 4.2, we
     looked to determine the faintest detectable magnitude as a
     function of angular separation from each parent star.  We began
     by sampling each of the final psf-subtracted and non-subtracted
     noise maps and selecting, for each pixel, the smaller of the two
     noise values.  The resulting composite noise map array therefore reflected the best
     sensitivities from each of the two final images.      Figure 2 displays a sample image sequence, where a psf-subtracted
     and non-subtracted image are combined to create a composite noise map.    

Once we had generated our composite noise maps, we declared an array of sample apparent K$_{s}$-magnitudes
     extending from 8 to 23 mags at intervals of 0.3 mags.  This
     selection included all potential brown dwarf magnitudes that we
     were likely to encounter.  We do note that some of the lowest
     luminosity brown dwarfs may have magnitudes dimmer than our
     23-magnitude limit.  However, since 23 magnitudes was effectively
     beyond even our most optimistic sensitivity estimates, we did not
     need to consider anything fainter than that.  We next transformed
     the apparent magnitudes to instrument counts using the parent
     star calibration data described in Section 3.1. 
  
     Returning to the  composite noise map, we determined the median values
     in a series of concentric 0$\arcsec$.20-thick rings centered on
     the noise map center.  The median values therefore
     represented typical noise as a function of distance from
     the central star.  For each noise value, we then determined the
     minimum apparent K$_{s}$-magnitude where signal exceeded the
     combined Poisson noise and ring noise by a factor greater or
     equal to 5.  In Figure 3 we plot resulting measurements for
     median survey sensitivities (middle
     curve), the best 10\% of
     observations (lower curve), and the worst 10\% of
     observations (top curve).  Refer to Table 3 for a summary of minimum detectable magnitudes for each of the individual targets.

     Another commonly used statistic for describing sensitivities for
     high-contrast companion surveys is the limiting differential
     magnitude as a function of angular separation from the parent
     star.  In other words, how many times dimmer may a companion
     object be before we lose it in the parent star noise?  Figure
     4
     plots differential magnitudes for median survey sensitivities  as
     well as the best
     and worst 10\% of observations.  

\subsection{Mass Sensitivities}
  Determining sensitivities according to companion mass is complicated
  by the fact that brown dwarfs of a given mass dim over time.
  Nonetheless, to get a general idea of detectable masses, we may
  assume different test ages and then use models by Burrows et
  al. (2001) or Baraffe et al. (2003) to
  transform our minimum detectable brightnesses into brown dwarf
  masses.   Figure
  5 shows a comparison of median sensitivities assuming 1 Gyr, solar
  age, and 10 Gyr target ages.   

\section{Results}
     After conducting all of our data analysis, we concluded that zero
     systems showed positive evidence of a brown dwarf companion.
     For Gliese 412 follow-up common proper motion observations, the
     available observing time was too short for us to positively
     confirm or reject common proper motion.  2MASS data tells us
     that, in the Gliese 412 neighborhood, the odds of our finding a
     field star in the PHARO field of view are about 1\%.  If it
     is a true companion, its magnitude would place it somewhere
     around an L9 dwarf classification.  In a survey of 80 target
     stars, a 1\% chance alignment is not particularly unusual, making
     a field star classification a reasonable potentiality.  In the
     end though, these speculations cannot confirm or reject the
     presence of a true brown dwarf companion.  At this point, we
     classify it as a non-brown dwarf detection until a time when we
     may confirm its substellar companion nature.  Table 2  presents
     our discovered field stars meeting the
     automated detection routine's sensitivity criteria. 

\section{Discussion}
   The observational data we have presented here clearly supports
    speculations of a ``brown dwarf desert'' at orbital separations
    comparable to our own outer solar system.  However, we emphasize
     that we cannot definitively assert that a brown dwarf desert exists before applying rigorous Monte Carlo
    simulations that take into account any observational biases.  For example, if the brown dwarf companion
    population were to have unusually high eccentricities, then the
    $\sim$100 AU projected separations that we believe we are
    investigating could in fact be representative of
    semi-major axes closer to 10 AU.  In that case, the 100 AU (true
    semi-major axis) brown
    dwarf companion population 
    could in fact be quite high since the members would spend the majority of
    their orbit outside of our field of view.  To address this
    issue, we conducted full-scale Monte Carlo simulations
    that account for the effects of differing orbital parameters.  In our
    upcoming paper (part II of this series) we discuss such population simulations at
    length.  One important early result though of such simulations is that approximate analytical solutions presented in McCarthy \&
    Zuckerman (2004) and Gizis et al. (2001), which assume  zero inclination and
    zero eccentricity, suffer from sytematic observational biases that cause them to
    dramatically understate their 
    uncertainties.  Thus, we caution the reader against firmly
    asserting a brown dwarf companion desert before reading our entire
    upcoming analysis.

\acknowledgments

We thank Jean Mueller, Rick Burruss, Karl Dunscombe, and the Palomar
Mountain crew for their support.  We also thank Sarah Higdon and James
Higdon for conducting observations for us.  We thank our anonymous
referee for their careful review and helpful suggestions.  J. C. C. and S. S. E. were
supported in part by NSF CAREER award AST-0328522.

Facilities: \facility{Palomar Observatory(PALAO/PHARO,WIRC)}.

\clearpage

\begin{figure}
\plotone{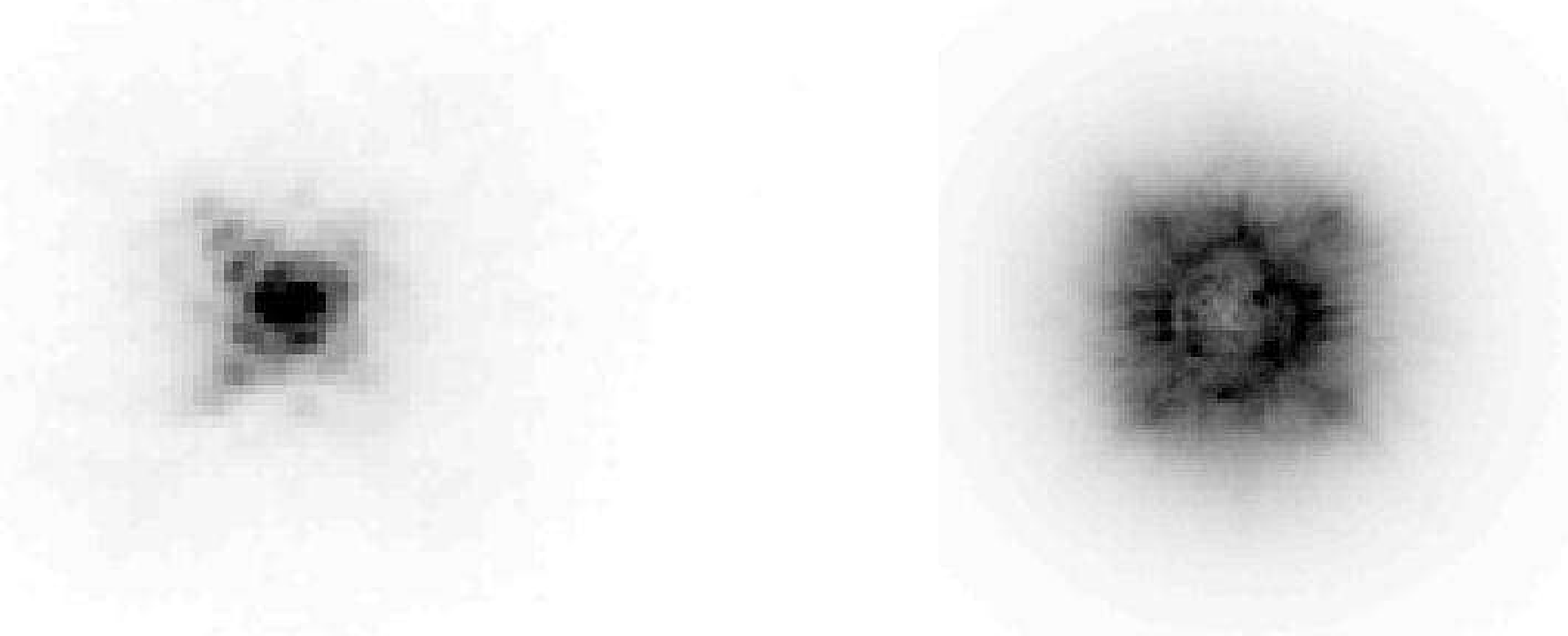}  
\caption{ Two reduced Gliese 183 images taken in December,
  2001 using the Palomar 200-inch Adaptive Optics and accompanying
  PHARO science camera.  The image on the left is a non-coronagraphed
  Gliese 183 image taken with a neutral density filter.  The image on
  the right was taken with no neutral density filter, but with a
  $0.\!''91$ spot positioned over the star.  The images illustrate
  PALAO's characteristic AO-reconstructed PSF as seen in both
  coronagraphic and non-coronagraphic imaging.} 
\end{figure}

\begin{figure}
\plotone{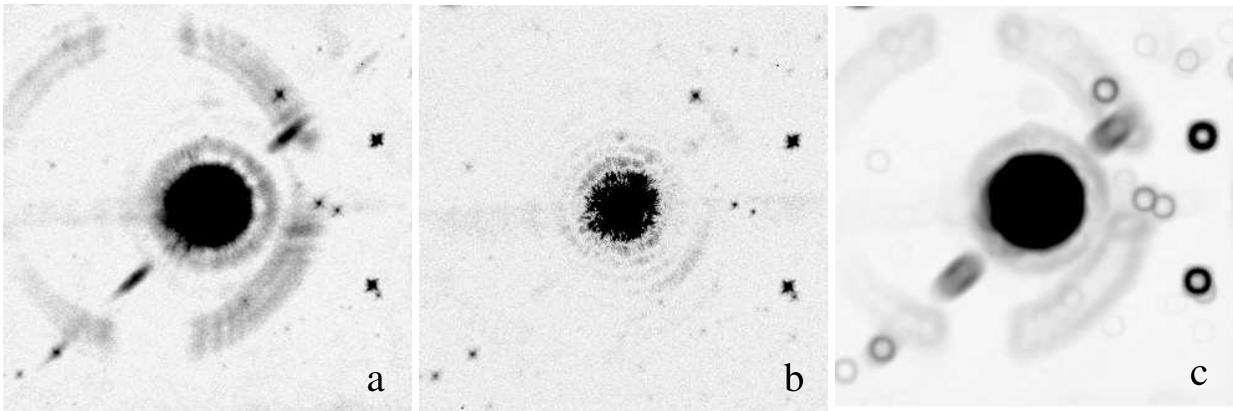} 
\caption{Gliese 740 Final Images and Noise Map.  a) Gliese 740
  Non-PSF-Subtracted Final Image.  b) Gliese 740 PSF-Subtracted Final
  Image.  c) A composite noise map generated by our automated
  search algorithm using images (a) and (b).  The
  doughnut-shaped features toward the outskirts in (c) represent
  high-noise regions caused by the detection algorithm's difficulty in
  identifying sources in close proximity to field stars.  The
  dark, slightly offset ring segments around the star in (a) and (c) represent
  internal instrument reflection.  The feature resembling
  an edge-on disk, extending to the upper right and lower left of the
  star in (a) and (c) is an artifact resulting from an oily smudge on one of the AO
  mirrors.  The smudge was identified and removed in late 2003.}
\end{figure}

\begin{figure}
\plotone{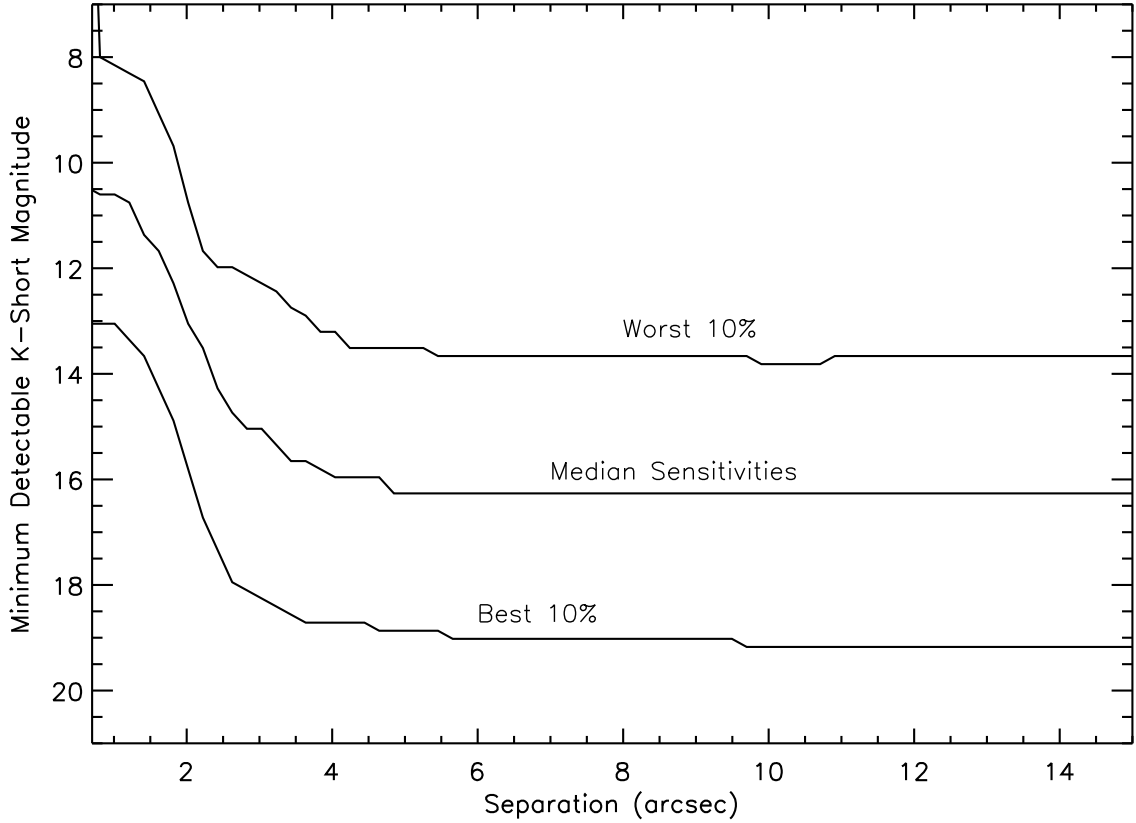}  
\caption{K$_{s}$-band sensitivity curves displaying limiting magnitude
  as a function of separation from the parent star.  The top curve
  represents the median sensitivities for the  worst 10\% of CHAOS
  observations. The middle curve represents median survey
  sensitivities.  The
  bottom curve represents median sensitivities for the best 10\% of
  observations.   ``Best 10\%'' and ``Worst 10\%'' are defined by a combination of parent star
  brightness, seeing conditions, and adaptive optics performance.  All minimum magnitudes correspond to 5-sigma detections.}
\end{figure}

\begin{figure}
\plotone{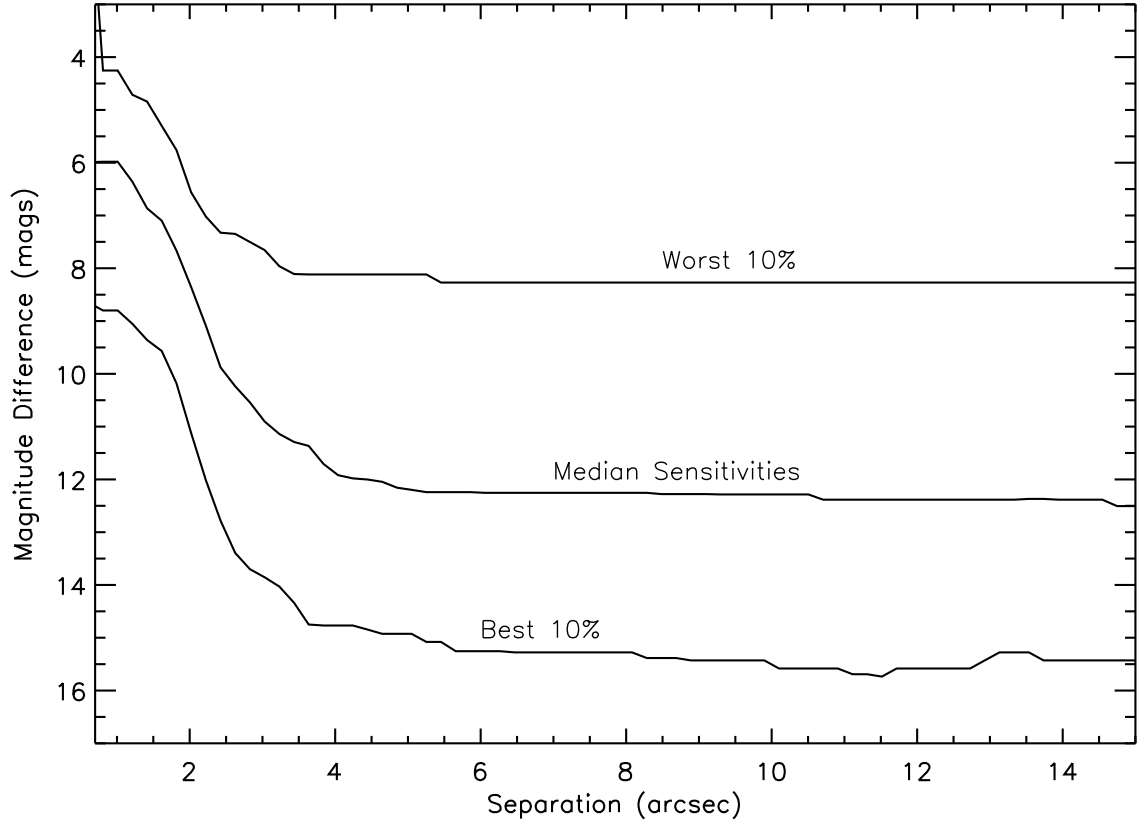} 
\caption{K$_{s}$-band sensitivity curves displaying limiting
  differential magnitude (K$_{s}$-companion minus K$_{s}$-parent) as a
  function of separation from the parent star.  The  middle curve
  represents median survey sensitivities.  The top curve represents
  median sensitivities for the worst 10\% of our data.  The
  bottom curve shows the best 10\%.  Limits represent 5-sigma detections.}
\end{figure} 

\begin{figure}
\plotone{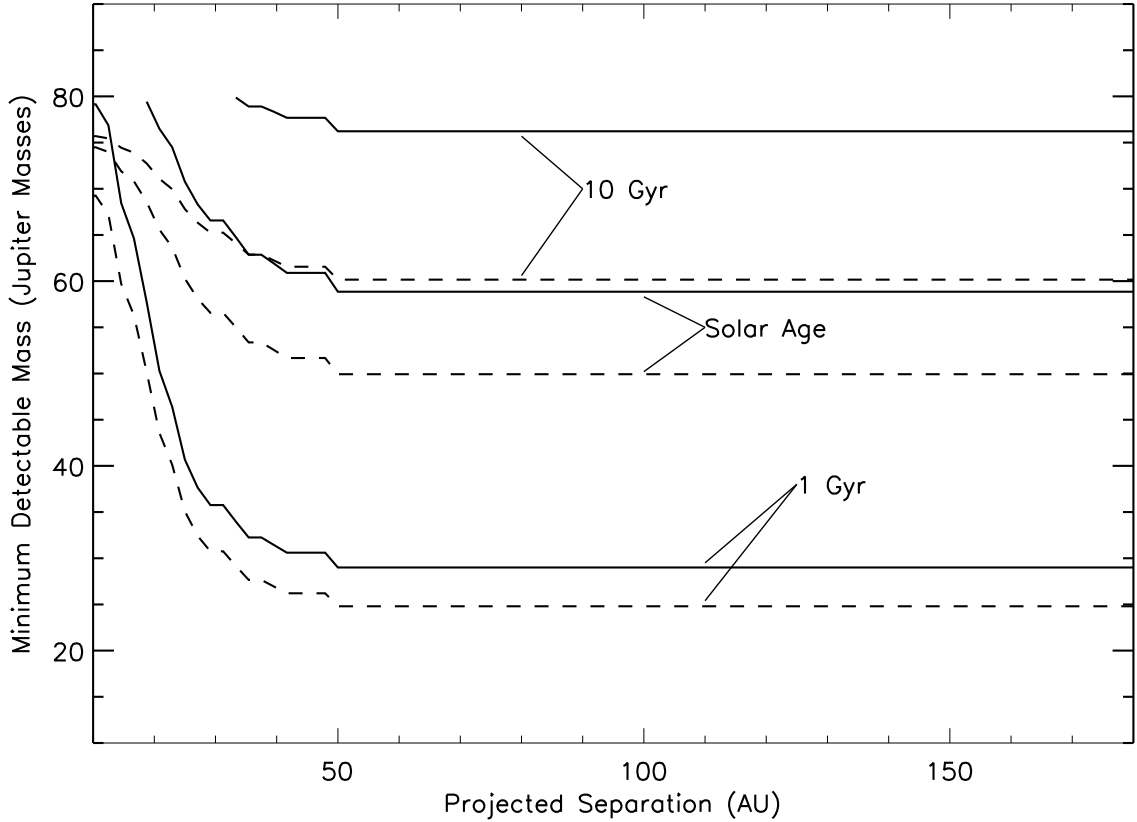}   
\caption{Minimum detectable mass as a function of projected separation
   for median target sensitivities and distances.  We compare
  results for assumed 1 Gyr, solar age, and 10 Gyr targets, using
  evolutionary models by Burrows et al. (2001) (solid curve) or
  Baraffe et al. (2003) (dashed curve).  We derived these
  curves by first plotting minimum detectable K-short magnitude
  versus arcsecond separation for all targets.  Next we
  median-combined the K-short curves to derive typical sensitivities (see Figure 3, middle curve).  We transformed these K-short magnitudes
  into masses using Hipparcos distances and Burrows et al. (2001)
  evolutionary models (solid curves) or Baraffe et al. (2003)
  evolutionary models (dashed curves).  Finally, we transformed our
  arcsecond axis to a projected AU separation using the median target distance.}     
\end{figure}

\begin{deluxetable}{lllllcc}
\tabletypesize{\scriptsize}
\rotate
\tablecaption{The CHAOS Target List}
\tablewidth{0pt}
\tablehead{
\colhead{Parallax} &    \colhead{\hspace{2cm} Proper Motion} & &
\colhead{V} & & Dates of Coronagraphic & Net Exposure \\
\colhead{(mas)} &
\colhead{RA (mas yr$^{-1}$)} & \colhead{Dec (mas yr$^{-1}$)} &
\colhead{(mag)} & \colhead{Name} & Observations & Time (sec) \\ 
}
\startdata    
549.01 &  \hspace{1.05cm}   \hspace{-0.2cm} -797.84 &       10326.93 &
9.54 &  Gliese 699 & 2002 Jun & 363 \\
392.40&    \hspace{1.05cm}   \hspace{-0.2cm} -580.20& \hspace{-0.2cm}
-4767.09&       7.49&   Gliese 411 & 2000 May & 218 \\
310.75&   \hspace{1.05cm}   \hspace{-0.2cm} -976.44&        17.97&
3.72&   Gliese 144 & 2000 Aug & 545 \\
280.27&    \hspace{1.05cm}  2888.92&        410.58& 8.09&   Gliese 15
& 2000 Aug & 482 \\
263.26&     \hspace{1.05cm}  571.27& \hspace{-0.2cm} -3694.25&
9.84&   Gliese 273 & 2000 Nov & 291 \\
206.94&     \hspace{1.05cm}  \hspace{-0.2cm} -4410.79&       943.32&
8.82&   Gliese 412 & 2002 Dec; 2004 Jul & 713 \\
205.22&     \hspace{1.05cm}  \hspace{-0.2cm} -1361.55&
\hspace{-0.2cm}   -505.00&   6.60&    Gliese 380 & 2001 Dec & 291 \\
204.60\tablenotemark{*} & \hspace{1.05cm} \hspace{-0.2cm} -505.00\tablenotemark{**} &
\hspace{-0.2cm} -62.00\tablenotemark{**} & \hspace{-0.1cm}
10.0\tablenotemark{***} & Gliese 388 & 2002 Dec & 291 \\
198.24& 
\hspace{1.05cm} \hspace{-0.2cm} -2239.33&       \hspace{-0.2cm}
-3419.86&       4.43&   Gliese 166 & 2001 Dec & 581 \\
194.44&      \hspace{1.05cm} 536.82& 385.54& 0.76&   Gliese 768 &
2002; 2004 Jul & 333 \\
184.13&   \hspace{1.05cm}    1778.46&        \hspace{-0.2cm} -1455.52&
8.46&   Gliese 526 & 2001 Jun & 654 \\
175.72&    \hspace{1.05cm}  82.86&  \hspace{-0.2cm} -3.67&  7.97&
Gliese 205 & 2001 Dec & 581 \\
174.23&    \hspace{1.05cm}  \hspace{-0.2cm} -829.34&
\hspace{-0.2cm} -878.81&        9.02&   Gliese 644 & 2002 Jun & 636 \\
173.41&    \hspace{1.05cm}   598.43& \hspace{-0.2cm} -1738.81&
4.67&   Gliese 764 & 2001 Jun & 799\\
170.26&    \hspace{1.05cm}  \hspace{-0.2cm} -578.86&
\hspace{-0.2cm} -1331.70&        9.12&   Gliese 752 & 2000 Sep; 2001
Jun & 1247\\
167.51&     \hspace{1.05cm}  995.12& \hspace{-0.2cm} -968.25&
8.98&   Gliese 908 & 2000 Aug & 654\\
153.24&     \hspace{1.05cm}  2074.37&        294.97& 5.57&   Gliese
892 & 2000 Sep; 2000 Nov & 1126\\
150.96&     \hspace{1.05cm}  \hspace{-0.2cm} -426.31&
\hspace{-0.2cm} -279.94&   \hspace{-0.1cm}     10.03&  Gliese 408 &
2002 Feb & 580\\
141.95&     \hspace{1.05cm}  1.08&   \hspace{-0.2cm} -774.24&
8.55&   Gliese 809 & 2000 Aug; 2004 Jul & 973\\
134.04&     \hspace{1.05cm}  758.04& \hspace{-0.2cm} -1141.22&
5.74&   Gliese 33 & 2000 Aug & 594\\
132.40&     \hspace{1.05cm}  3421.44&        \hspace{-0.2cm} -1599.27&
5.17&   Gliese 53 & 2000 Aug & 908\\
131.12&     \hspace{1.05cm}  1128.00&   \hspace{-0.2cm} -1074.30&
9.05&   Gliese 514 & 2000 May; 2001 Jun & 1199\\
129.54&     \hspace{1.05cm}  \hspace{-0.2cm} -580.47&
\hspace{-0.2cm} -1184.81&       7.54&   Gliese 673 & 2000 Sep; 2001
Jun & 1145\\
119.46&    \hspace{1.05cm}   \hspace{-0.2cm} -705.06&        292.93&
4.24&   Gliese 475 & 2001 Jun & 872\\
119.05&    \hspace{1.05cm}
\hspace{-0.2cm} -291.42&        \hspace{-0.2cm} -750.00&   3.42&
Gliese 695 & 2002 Jun & 581 \\
116.92&    \hspace{1.05cm}  \hspace{-0.2cm} -271.97&        254.93&
9.76&   Gliese 450 & 2002 Feb; 2002 Dec & 1708\\
115.43&    \hspace{1.05cm}   \hspace{-0.2cm} -163.17&
\hspace{-0.2cm} -98.92& 4.39&   Gliese 222 & 2000 Nov & 273\\
113.46&    \hspace{1.05cm}  550.74& \hspace{-0.2cm} -1109.30&
6.22&   Gliese 183 & 2000 Nov; 2001 Dec & 1163\\
109.95&    \hspace{1.05cm}   \hspace{-0.2cm} -2946.70&        184.52&
9.31&   Gliese 424 & 2001 Jun & 618\\
109.23&    \hspace{1.05cm}  \hspace{-0.2cm} -801.94&        882.70&
4.23&   Gliese 502 & 2002 Feb & 545\\
109.23&     \hspace{1.05cm}  124.03& 1.30& \hspace{-0.1cm}  10.08&
GJ2066 & 2001 Dec & 600\\
109.21&     \hspace{1.05cm}  4003.69&        \hspace{-0.2cm} -5813.00&
6.42&   Gliese 451 & 2002 Feb; 2002 Jun & 1708\\
104.81&    \hspace{1.05cm}   \hspace{-0.2cm} -13.95& \hspace{-0.2cm}
-380.46&        5.31&   Gliese 434 & 2000 May & 654\\
102.35&     \hspace{1.05cm}  \hspace{-0.2cm} -39.18& 383.41& 8.10&
Gliese 638 & 2000 May & 273\\
102.27&    \hspace{1.05cm}  455.22& \hspace{-0.2cm} -307.63&
5.77&   Gliese 631 & 2000 May & 654\\
100.24&    \hspace{1.05cm}   582.05& \hspace{-0.2cm} -246.83&
5.63&   Gliese 75 & 2000 Nov & 297\\
99.44&     \hspace{1.05cm}  730.10&  89.27&  9.56&   Gliese 49 & 2000
Aug & 908\\
98.97&     \hspace{1.05cm}  \hspace{-0.2cm} -32.32& 216.48& 6.49&
HIP 68184\tablenotemark{****} & 2000 May & 864\\
98.26&    \hspace{1.05cm}   \hspace{-0.2cm} -824.78&        599.52&
9.71&   Gliese 536 & 2000 May & 320\\
98.12&     \hspace{1.05cm}  305.20&  \hspace{-0.2cm} -475.59&
8.62&   Gliese 172 & 2000 Nov; 2001 Dec & 1160\\
96.98&    \hspace{1.05cm}   \hspace{-0.2cm} -455.08&
\hspace{-0.2cm} -280.37&        9.17&   Gliese 846 & 2000 Aug & 900\\
94.93&    \hspace{1.05cm}  1262.29&        \hspace{-0.2cm} -91.53&
4.05&   Gliese 124 & 2000 Nov; 2002 Dec & 872\\
93.81&    \hspace{1.05cm}  \hspace{-0.2cm} -6.56&  \hspace{-0.2cm}
-5.69&  10.2&   HIP 109119\tablenotemark{****} & 2002 Jun & 900\\
93.79&     \hspace{1.05cm}  \hspace{-0.2cm} -497.89&        85.88&
8.61&   Gliese 617 & 2002 Jun & 1090\\
93.36&     \hspace{1.05cm}   \hspace{-0.2cm} -179.67&
\hspace{-0.2cm} -98.24& 6.53&   Gliese 688 & 2002 Jun; 2004 Jul & 727\\
92.98&     \hspace{1.05cm}  \hspace{-0.2cm} -916.86&
\hspace{-0.2cm} -1137.91&       7.70&    Gliese 653 & 2000 May & 981\\
92.75&     \hspace{1.05cm}  \hspace{-0.2cm} -28.77& \hspace{-0.2cm}
-397.85&        8.49&   Gliese 488 & 2000 May & 899\\
92.20&     \hspace{1.05cm}  1151.61&        \hspace{-0.2cm} -246.32&
4.84&   Gliese 92 & 2000 Sep; 2000 Nov & 545\\
91.74&     \hspace{1.05cm}  740.96& \hspace{-0.2cm} -271.18&
3.59&   Gliese 449 & 2002 Feb & 818\\
90.11&    \hspace{1.05cm}   \hspace{-0.2cm} -316.17&
\hspace{-0.2cm} -468.33&        6.38&   Gliese 706 & 2002 Jun & 818\\
90.03&     \hspace{1.05cm}  \hspace{-0.2cm} -461.07&
\hspace{-0.2cm} -370.88&        5.88&   Gliese 27 & 2000 Aug & 908\\
90.02&    \hspace{1.05cm}   \hspace{-0.2cm} -194.47&
\hspace{-0.2cm} -1221.78&       9.22&   Gliese 740 & 2000 May; 2000
Aug & 1200\\
89.92&     \hspace{1.05cm}  311.20&  \hspace{-0.2cm} -1282.17&
3.85&   Gliese 603 & 2002 Feb & 872\\
89.70&     \hspace{1.05cm}  \hspace{-0.2cm} -921.19&
\hspace{-0.2cm} -1128.23&       10.08&  Gliese 654 & 2002 Jun & 581\\
88.17&     \hspace{1.05cm}   \hspace{-0.2cm} -60.95& \hspace{-0.2cm}
-358.10& 2.68&   Gliese 534 & 2002 Feb; 2002 Jun & 854\\
87.17&    \hspace{1.05cm}   \hspace{-0.2cm} -65.14& 175.55& 8.30&
Gliese 169 & 2000 Nov & 1090\\
86.69&    \hspace{1.05cm}    \hspace{-0.2cm} -442.75&        216.84&
6.00&   Gliese 567 & 2002 Feb & 872\\
85.48&       \hspace{1.05cm}  241.12&  370.51& 9.15& Gliese 572 & 2000
May & 727\\
85.08&    \hspace{1.05cm}   \hspace{-0.2cm} -225.51&
\hspace{-0.2cm} -68.52& 4.42&   Gliese 598 & 2002 May & 1163\\
85.06&    \hspace{1.05cm}   296.73& 26.93&  3.77&   Gliese 848 & 2000
Nov & 1090 \\
83.85&     \hspace{1.05cm}   592.12& \hspace{-0.2cm} -197.10& 8.31&
HIP54646\tablenotemark{****} & 2000 May & 928\\
81.69&     \hspace{1.05cm}  2.70&    \hspace{-0.2cm} -523.61&
6.21&   Gliese 211 & 2001 Dec & 1090\\
80.13&     \hspace{1.05cm}   3.16&   \hspace{-0.2cm} -517.26&
9.78&   Gliese 212 & 2002 Dec & 1163\\
80.07&     \hspace{1.05cm}  \hspace{-0.2cm} -689.12&        120.68&
10.15&  Gliese 390 & 2002 Feb & 1163\\
79.80&    \hspace{1.05cm}   \hspace{-0.2cm} -485.48&
\hspace{-0.2cm} -234.40& 5.96&   Gliese 324 & 2001 Dec & 2616\\ 
78.87&    \hspace{1.05cm}  \hspace{-0.2cm} -504.10& 110.00&    7.20&
Gliese 349 & 2001 Dec; 2002 Dec & 2117\\
78.14&    \hspace{1.05cm}   \hspace{-0.2cm} -531.00&   3.65&   6.44&
Gliese 675 & 2000 May & 1090\\
78.07&   \hspace{1.05cm}    537.93& 140.05& 10.36&  Gliese 863 & 2000
Aug & 1200\\
77.82&     \hspace{1.05cm}   \hspace{-0.2cm} -177.02&
\hspace{-0.2cm} -33.45& 4.82&   Gliese 395 & 2002 Feb & 1163\\
76.26&     \hspace{1.05cm}   \hspace{-0.2cm} -92.28& 120.76& 7.46&
Gliese 775 & 2000 May & 981\\
74.45&    \hspace{1.05cm}   \hspace{-0.2cm} -394.18&
\hspace{-0.2cm} -581.75&        8.42&   Gliese 69 & 2000 Nov & 1226\\
73.58&    \hspace{1.05cm}   8.61&   \hspace{-0.2cm} -248.72&
10.25&  NN3371 & 2000 Nov & 1199\\
69.73&     \hspace{1.05cm}  86.08&  817.89& 3.41&   Gliese 807 & 2002
Jun & 1163\\
68.63&     \hspace{1.05cm}   \hspace{-0.2cm} -236.06&
\hspace{-0.2cm} -399.07&        4.04&  Gliese 549 & 2002 Feb & 1090\\
60.80&     \hspace{1.05cm}  \hspace{-0.2cm} -126.64&
\hspace{-0.2cm} -172.00&   5.31&   Gliese 678 & 2002 Jun & 872\\
58.50&     \hspace{1.05cm}  \hspace{-0.2cm} -413.14&        30.44&
6.01&   Gliese 327 & 2001 Dec & 1272\\
54.26&     \hspace{1.05cm}   222.07& \hspace{-0.2cm} -92.62 &
5.08    & Gliese 368 & 2002 Dec & 1163\\
53.85&    \hspace{1.05cm}  \hspace{-0.2cm} -68.45& 169.72  & 4.80   &
Gliese 41  & 2001 Dec & 2326\\
50.71&    \hspace{1.05cm}   \hspace{-0.2cm} -6.05&  \hspace{-0.2cm}
-16.03  & 6.17  & HIP113421\tablenotemark{****} & 2002 Sep & 1272\\
45.43&    \hspace{1.05cm}  \hspace{-0.2cm} -49.04& \hspace{-0.2cm}
-573.07 & 8.03  & Gliese 31.4 & 2001 Dec & 1744\\
\enddata
\tablecomments{Parallax, proper motion, and V-magnitude are from Hipparcos (Perryman et al. 1997) unless otherwise noted.  All names follow the Gliese catalogue system (Gliese \& Jahreiss 1995) unless otherwise noted.}                                
 
\tablenotetext{*}{Yale Trigonometric Parallaxes (van Altena, Lee, \& Hoffleit 2001)}                                                                    
\tablenotetext{**}{AGK3 Catalogue (Bucciarelli et al. 1996)}            
\tablenotetext{***}{Brorfelde Meridian Catalogues (Laustsen 1996)}     
\tablenotetext{****}{Hipparcos (Perryman et al. 1997)} 
\end{deluxetable}

\begin{deluxetable}{lllll}
\tabletypesize{\scriptsize}
\tablecaption{Discovered Field Stars}
\tablewidth{0pt}
\tablehead{
 & & & \colhead{Observation} & \\
\colhead{Parent Field}  &\colhead{$\rho$ (arcsec)\tablenotemark{a}}
&\colhead{$\theta$ (deg)\tablenotemark{b}} &\colhead{Date}
&\colhead{K$_{s}$ (mag)}  
}
\startdata
Gliese 809 & 10.62 $\pm$ 0.12 & 89.4 $\pm$ 0.4 & 2000 Aug & 14.4 $\pm$
0.4 \\
Gliese 49 & 13.71 $\pm$ 0.14 & 230.7 $\pm$ 0.3 & 2000 Aug & 11.4 $\pm$
0.3 \\
Gliese 740 & 16.35 $\pm$ 0.16 & 51.1 $\pm$ 0.3 & 2000 May & 11.2 $\pm$
0.3 \\
Gliese 740 & 10.50 $\pm$ 0.12 & 107.7 $\pm$ 0.6 & 2000 May & 12.5 $\pm$
0.5 \\
Gliese 740 & 9.29 $\pm$ 0.11 & 330.3 $\pm$ 0.6 & 2000 May & 13.0 $\pm$
0.4 \\
Gliese 740 & 13.63 $\pm$ 0.15 & 310.1 $\pm$ 0.4 & 2000 May & 12.8 $\pm$ 0.4 \\
Gliese 740 & 11.69 $\pm$ 0.12 & 298.6 $\pm$ 0.3 & 2000 May & 9.4
$\pm$ 0.3 \\
Gliese 740 & 8.27 $\pm$ 0.08 & 274.8 $\pm$ 0.3 & 2000 May & 10.8
$\pm$ 0.3 \\
Gliese 740 & 7.06 $\pm$ 0.07 & 271.4 $\pm$ 0.3 & 2000 May & 10.7
$\pm$ 0.3 \\
Gliese 740 & 11.64 $\pm$ 0.12 & 251.3 $\pm$ 0.3 & 2000 May & 9.1
$\pm$ 0.3 \\
Gliese 740 & 8.49 $\pm$ 0.09 & 215.1 $\pm$ 0.3 & 2000 May & 10.3
$\pm$ 0.3 \\
Gliese 740 & 12.35 $\pm$ 0.14 & 218.0 $\pm$ 0.4 & 2000 May & 12.4 $\pm$
0.4 \\
Gliese 740 & 16.63 $\pm$ 0.17 & 231.4 $\pm$ 0.4 & 2000 May & 12.3 $\pm$
0.3 \\
Gliese 740 & 4.44 $\pm$ 0.05 & 179.1 $\pm$ 0.3 & 2000 May & 11.8 $\pm$
0.3 \\
Gliese 75 & 10.22 $\pm$ 0.10 & 328.5 $\pm$ 0.3 & 2000 Nov & 10.7 $\pm$
0.3 \\
Gliese 172 & 11.04 $\pm$ 0.11 & 99.7 $\pm$ 0.3 & 2000 Nov & 10.7 $\pm$
1.1 \\
Gliese 69 & 8.62 $\pm$ 0.09 & 110.8 $\pm$ 0.3 & 2000 Nov & 12.9 $\pm$
0.3 \\
Gliese 69 & 6.90 $\pm$ 0.09 & 320.7 $\pm$ 0.3 & 2000 Nov & 12.3 $\pm$
0.4 \\
Gliese 892 & 13.81 $\pm$ 0.14 & 63.6 $\pm$ 0.3 & 2000 Nov & 11.5 $\pm$
1.1 \\
Gliese 892 & 11.23 $\pm$ 0.11 & 21.6 $\pm$ 0.3 & 2000 Nov & 13.2 $\pm$
1.1 \\
Gliese 752 & 4.79 $\pm$ 0.05 & 84.5 $\pm$ 0.3 & 2000 Sep & 12.5 $\pm$
0.2 \\
Gliese 752 & 8.38 $\pm$ 0.08 & 76.7 $\pm$ 0.3 & 2000 Sep & 12.3 $\pm$
0.2 \\
Gliese 752 & 7.91 $\pm$ 0.08 & 68.2 $\pm$ 0.3 & 2000 Sep & 11.8 $\pm$
0.2 \\
Gliese 752 & 12.18 $\pm$ 0.12 & 32.2 $\pm$ 0.3 & 2000 Sep & 13.1 $\pm$
0.3 \\
Gliese 752 & 9.81 $\pm$ 0.10 & 359.9 $\pm$ 0.3 & 2000 Sep & 13.3 $\pm$
0.3 \\
Gliese 752 & 5.92 $\pm$ 0.06 & 340.5 $\pm$ 0.3 & 2000 Sep & 12.2 $\pm$
0.2 \\
Gliese 752 & 12.74 $\pm$ 0.13 & 342.7 $\pm$ 0.3 & 2000 Sep & 15.0 $\pm$
0.5 \\
Gliese 752 & 6.60 $\pm$ 0.07 & 320.3 $\pm$ 0.3 & 2000 Sep & 11.1 $\pm$
0.2 \\
Gliese 752 & 11.15 $\pm$ 0.12 & 335.0 $\pm$ 0.4 & 2000 Sep & 15.0 $\pm$
0.4 \\
Gliese 752 & 7.22 $\pm$ 0.07 & 312.2 $\pm$ 0.3 & 2000 Sep & 11.8 $\pm$
0.2 \\
Gliese 752 & 8.88 $\pm$ 0.09 & 318.5 $\pm$ 0.3 & 2000 Sep & 12.1 $\pm$
0.2 \\
Gliese 752 & 10.38 $\pm$ 0.11 & 314.4 $\pm$ 0.3 & 2000 Sep & 15.4 $\pm$
0.6 \\
Gliese 752 & 13.39 $\pm$ 0.13 & 326.1 $\pm$ 0.3 & 2000 Sep & 12.3
$\pm$ 0.2 \\
Gliese 752 & 10.58 $\pm$ 0.11 & 304.0 $\pm$ 0.3 & 2000 Sep & 14.6
$\pm$ 0.3 \\
Gliese 752 & 9.11 $\pm$ 0.09 & 284.8 $\pm$ 0.3 & 2000 Sep & 5.8
$\pm$ 0.2 \\
Gliese 752 & 16.74 $\pm$ 0.17 & 327.1 $\pm$ 0.3 & 2000 Sep & 13.0
$\pm$ 0.3 \\
Gliese 752 & 11.6 $\pm$ 0.12 & 267.0 $\pm$ 0.3 & 2000 Sep & 14.8
$\pm$ 0.4 \\
Gliese 752 & 9.02 $\pm$ 0.10 & 259.1 $\pm$ 0.4 & 2000 Sep & 15.0
$\pm$ 0.3 \\
Gliese 752 & 6.39 $\pm$ 0.06 & 209.2 $\pm$ 0.3 & 2000 Sep & 12.1
$\pm$ 0.2 \\
Gliese 752 & 10.00 $\pm$ 0.10 & 207.6 $\pm$ 0.3 & 2000 Sep & 12.5
$\pm$ 0.2 \\
Gliese 752 & 7.94 $\pm$ 0.08 & 163.0 $\pm$ 0.3 & 2000 Sep & 12.5
$\pm$ 0.2 \\
Gliese 752 & 7.48 $\pm$ 0.08 & 125.0 $\pm$ 0.3 & 2000 Sep & 13.1 $\pm$
0.3 \\
Gliese 673 & 7.29 $\pm$ 0.07 & 269.3 $\pm$ 0.3 & 2000 Sep & 9.5 $\pm$
0.2 \\
Gliese 41 & 13.54 $\pm$ 0.14 & 38.3 $\pm$ 0.3 & 2001 Dec & 9.7 $\pm$
0.3 \\
Gliese 412 & 7.31 $\pm$ 0.07 & 167.9 $\pm$ 0.3 & 2002 Dec & 7.5 $\pm$
0.2 \\
Gliese 678 & 13.74 $\pm$ 0.14 & 241.1 $\pm$ 0.3 & 2002 Jun & 10.6 $\pm$
0.3 \\
Gliese 678 & 16.86 $\pm$ 0.17 & 317.0 $\pm$ 0.3 & 2002 Jun & 5.5 $\pm$
0.3 \\
Gliese 688 & 12.52 $\pm$ 0.13 & 299.9 $\pm$ 0.3 & 2002 Jun & 13.0 $\pm$
0.3 \\

\enddata
\tablecomments{All detections represent $\geq$5-sigma signal-to-noise levels.}
\tablenotetext{a}{Separation from central star.}
\tablenotetext{b}{Position angle, measured counter-clockwise from
  central star's north-south axis.}

\end{deluxetable}

\begin{deluxetable}{lcllll}
\tabletypesize{\scriptsize}
\tablecaption{CHAOS Target Sensitivities}
\tablewidth{0pt}
\tablehead{
\colhead{}  &\colhead{} &\colhead{} &\colhead{} &\colhead{} &
\colhead{\hspace{-4.6cm}Faintest Detectable Apparent} \\
\colhead{}  &\colhead{} &\colhead{} &\colhead{} &\colhead{} &
\colhead{\hspace{-4.6cm}$K_{s}$-magnitude by Separation} \\
\colhead{Target Name} & \colhead{\hspace{0.2cm} $0.\!''$96} & \colhead{$1.\!''$92} & \colhead{$3.\!''$04} & \colhead{$4.\!''$96} \\
}
\startdata
Gliese 53 & \hspace{0.2cm} 8.9 & 12.3 & 14.7 & 15.7 \\
HIP 109119 & \hspace{0.2cm} 13.2 & 15.0 & 16.6 & 16.6 \\
Gliese 809 & \hspace{0.2cm} 10.8 & 13.8 & 16.6 & 17.2 \\
Gliese 15 & \hspace{0.2cm} 12.6 & 15.0 & 18.1 & 19.0 \\
HIP 111313 & \hspace{0.2cm} 15.0 & 16.3 & 16.3 & 16.3 \\
Gliese 846 & \hspace{0.2cm} 12.3 & 14.4 & 16.6 & 16.6 \\
Gliese 27 & \hspace{0.2cm} 10.4 & 11.0 & 15.0 & 16.6 \\
Gliese 144 & \hspace{0.2cm} 11.4 & 13.5 & 16.3 & 18.1 \\
Gliese 49 & \hspace{0.2cm} 11.4 & 12.9 & 15.0 & 16.9 \\
Gliese 908 & \hspace{0.2cm} 17.5 & 17.5 & 17.8 & 17.8 \\
Gliese 33 & \hspace{0.2cm} 11.4 & 13.2 & 16.3 & 17.5 \\
HIP 73470 & \hspace{0.2cm} 9.2 & 13.5 & 14.4 & 15.3 \\
HIP 54646 & \hspace{0.2cm} 11.1 & 13.2 & 15.7 & 16.3 \\
HIP 68184 & \hspace{0.2cm} 11.4 & 13.2 & 15.0 & 16.9 \\
HIP 98698 & \hspace{0.2cm} 15.0 & 15.0 & 16.3 & 16.3 \\
Gliese 638 & \hspace{0.2cm} 12.6 & 14.4 & 16.6 & 16.9 \\
Gliese 411 & \hspace{0.2cm} 12.3 & 13.8 & 17.2 & 18.1 \\
Gliese 434 & \hspace{0.2cm} 9.5 & 11.7 & 14.7 & 15.7 \\
Gliese 653 & \hspace{0.2cm} 8.0 & 10.4 & 12.6 & 13.8 \\
Gliese 631 & \hspace{0.2cm} 8.0 & 10.1 & 13.2 & 15.0 \\
Gliese 488 & \hspace{0.2cm} 9.2 & 11.1 & 14.1 & 15.3 \\
Gliese 536 & \hspace{0.2cm} 8.6 & 11.0 & 14.4 & 15.7 \\
Gliese 675 & \hspace{0.2cm} 8.9 & 10.8 & 13.5 & 15.3 \\
Gliese 740 & \hspace{0.2cm} 8.6 & 11.1 & 13.8 & 15.7 \\
Gliese 514 & \hspace{0.2cm} 12.3 & 14.1 & 16.9 & 18.1 \\
Gliese 75 & \hspace{0.2cm} 9.5 & 12.3 & 14.4 & 16.3 \\
Gliese 172 & \hspace{0.2cm} 10.4 & 13.5 & 15.0 & 16.6 \\
NN3371 & \hspace{0.2cm} 11.7 & 13.8 & 15.0 & 16.3 \\
Gliese 273 & \hspace{0.2cm} 13.5 & 16.9 & 18.7 & 19.0 \\
Gliese 222 & \hspace{0.2cm} 8.9 & 11.7 & 15.3 & 16.3 \\
Gliese 183 & \hspace{0.2cm} 15.3 & 15.3 & 15.7 & 15.0 \\
Gliese 124 & \hspace{0.2cm} 8.0 & 8.0 & 11.1 & 13.2 \\
Gliese 212 & \hspace{0.2cm} 11.0 & 13.5 & 13.5 & 13.5 \\
Gliese 169 & \hspace{0.2cm} 9.2 & 10.8 & 14.1 & 15.0 \\
Gliese 848 & \hspace{0.2cm} 15.0 & 15.7 & 15.0 & 15.0 \\
Gliese 69 & \hspace{0.2cm} 9.8 & 11.4 & 14.4 & 16.3 \\
Gliese 92 & \hspace{0.2cm} 8.0 & 9.8 & 12.9 & 15.0 \\
Gliese 892 & \hspace{0.2cm} 11.0 & 15.3 & 17.8 & 18.1 \\
Gliese 752 & \hspace{0.2cm} 11.4 & 13.8 & 16.6 & 17.2 \\
Gliese 673 & \hspace{0.2cm} 10.4 & 13.2 & 15.0 & 16.3 \\
Gliese 41 & \hspace{0.2cm} 15.3 & 15.3 & 15.3 & 15.3 \\
Gliese 324 & \hspace{0.2cm} 8.9 & 11.0 & 13.8 & 15.0 \\
Gliese 380 & \hspace{0.2cm} 13.8 & 16.6 & 17.5 & 17.5 \\
Gliese 211 & \hspace{0.2cm} 13.8 & 14.1 & 14.7 & 15.3 \\
Gliese 166 & \hspace{0.2cm} 16.6 & 16.3 & 16.3 & 16.3 \\
Gliese 31.4 & \hspace{0.2cm} 13.5 & 13.5 & 13.5 & 13.8 \\
Gliese 349 & \hspace{0.2cm} 11.1 & 13.2 & 15.7 & 15.0 \\
Gliese 205 & \hspace{0.2cm} 16.6 & 16.6 & 16.6 & 17.2 \\
GJ2066 &  \hspace{0.2cm}  15.7 & 15.0 & 15.0 & 16.6 \\
Gliese 327 & \hspace{0.2cm} 14.4 & 15.0 & 15.0 & 15.3 \\
Gliese 526 & \hspace{0.2cm} 14.4 & 16.9 & 18.7 & 18.7 \\
Gliese 475 & \hspace{0.2cm} 12.9 & 15.0 & 17.2 & 17.5 \\
Gliese 424 & \hspace{0.2cm} 13.8 & 15.0 & 18.7 & 19.0 \\
Gliese 764 & \hspace{0.2cm} 10.8 & 13.5 & 16.6 & 18.1 \\
Gliese 388 & \hspace{0.2cm} 8.3 & 10.4 & 11.1 & 11.1 \\
Gliese 368 & \hspace{0.2cm} 11.1 & 13.2 & 13.8 & 13.8 \\
Gliese 412 & \hspace{0.2cm} 11.4 & 11.7 & 11.7 & 11.7 \\
Gliese 395 & \hspace{0.2cm} 14.7 & 15.0 & 15.7 & 15.0 \\
Gliese 451 & \hspace{0.2cm} 17.5 & 18.1 & 18.4 & 18.7 \\
Gliese 450 & \hspace{0.2cm} 12.9 & 14.4 & 14.4 & 14.4 \\
Gliese 534 & \hspace{0.2cm} 12.6 & 12.6 & 13.2 & 13.2 \\
Gliese 502 & \hspace{0.2cm} 9.5 & 11.4 & 13.8 & 15.7 \\
Gliese 549 & \hspace{0.2cm} 13.8 & 14.4 & 14.7 & 15.0 \\
Gliese 390 & \hspace{0.2cm} 12.3 & 14.4 & 15.0 & 15.0 \\
Gliese 449 & \hspace{0.2cm} 8.0 & 10.4 & 12.9 & 14.4 \\
Gliese 408 & \hspace{0.2cm} 17.8 & 17.8 & 17.8 & 17.8 \\
Gliese 567 & \hspace{0.2cm} 16.6 & 16.6 & 16.6 & 16.6 \\
Gliese 603 & \hspace{0.2cm} 10.4 & 12.9 & 15.0 & 16.3 \\
Gliese 644 & \hspace{0.2cm} 10.4 & 14.1 & 15.0 & 16.9 \\
Gliese 678 & \hspace{0.2cm} 14.1 & 14.4 & 14.4 & 14.4 \\
Gliese 768 & \hspace{0.2cm} 8.0 & 8.6 & 11.0 & 14.1 \\
Gliese 807 & \hspace{0.2cm} 8.0 & 9.8 & 11.0 & 13.8 \\
Gliese 706 & \hspace{0.2cm} 11.4 & 14.4 & 16.3 & 17.2 \\
Gliese 699 & \hspace{0.2cm} 14.7 & 17.2 & 19.0 & 20.2 \\
Gliese 617 & \hspace{0.2cm} 11.0 & 15.0 & 16.6 & 17.2 \\
Gliese 688 & \hspace{0.2cm} 15.3 & 15.3 & 15.7 & 16.3 \\
Gliese 695 & \hspace{0.2cm} 8.6 & 10.8 & 13.2 & 15.0 \\
Gliese 654 & \hspace{0.2cm} 15.7 & 15.7 & 15.7 & 15.0 \\
Gliese 598 & \hspace{0.2cm} 8.3 & 11.0 & 14.1 & 15.7 \\
HIP 113421 & \hspace{0.2cm} 13.2 & 13.2 & 13.2 & 13.5 \\
\enddata
\end{deluxetable}

\end{document}